\documentclass[10pt, preprint]{aastex}

\usepackage{graphicx}
\usepackage{epsfig}
\newcommand{\vel}{ {\bf v} }

\newcommand{\zhat}{{\bf {\hat e}}_z}
\newcommand{\bB}{{\bf B}}
\newcommand{ \bozx }{ B_{0}(x)}
\newcommand{ \ez }{ {\rm\bf e}_z}
\newcommand{ \ex }{ {\rm\bf e}_x}

\newcommand{ \curl }{ {\mbox{\boldmath$\times$}} }
\newcommand{ \dotp }{ {\mbox{\boldmath$\cdot$}} }
\newcommand{ \bnabla }{ {\mbox{\boldmath$\nabla$}} }
\newcommand{ \ekx }{ e^{i[kz - \omega t]} }

\newcommand{ \db }{ \partial_x \left(\frac{B_0^2}{2}\right) }
\newcommand{ \ddb }{ \partial^2_x \left(\frac{B_0^2}{2}\right) }

\begin{document}
\title{SEISMIC HALOS AROUND ACTIVE REGIONS: AN MHD THEORY}
\author{Shravan M. Hanasoge}
\affil{W. W. Hansen Experimental Physics Laboratory, Stanford University, Stanford, CA 94305}

\email{shravan@stanford.edu}
\begin{abstract}
Comprehending the manner in which magnetic fields affect propagating waves is a first step toward constructing accurate helioseismic models of active region sub-surface structure and dynamics. Here, we present a numerical method to compute the linear interaction of waves with magnetic fields embedded in a solar-like stratified background. The ideal Magneto-Hydrodynamic (MHD) equations are solved in a 3-dimensional box that straddles the solar photosphere, extending from 35 Mm within to 1.2 Mm into the atmosphere. One of the challenges in performing these simulations involves generating a Magneto-Hydro-Static (MHS) state wherein the stratification assumes horizontal inhomogeneity in addition to the strong vertical stratification associated with the near-surface layers. Keeping in mind that the aim of this effort is to understand and characterize linear MHD interactions, we discuss a means of computing statically consistent background states. Power maps computed from simulations of waves interacting with thick flux tubes of peak photospheric field strengths 600 G and 3000 G are presented. Strong modal power reduction in the `umbral' regions of the flux tube enveloped by a halo of increased wave power are seen in the simulations with the thick flux tubes. These enhancements are also seen in Doppler velocity power maps of active regions observed in the Sun, leading us to propose that the halo has MHD underpinnings.
\end{abstract}
\keywords{Sun: helioseismology---Sun: interior---Sun: oscillations---waves---hydrodynamics}

\section{INTRODUCTION}
The complexity of the solar background state, subtleties in the dynamics of wave propagation in the near-surface layers, and the inherently anisotropic, tensorial nature of magnetic fields disadvantage analytically driven MHD studies. There have been many theoretical efforts to model MHD interactions in flux concentrations but have proven to be somewhat restrictive in the scope of problems addressed given the effort required to construct these models. In this regard, numerical forward modeling of wave propagation \citep[e.g.][]{hanasoge1,hanasoge3,cameron07,shelyag,khomenko,parchevsky} has been relatively successful at making sense of the sometimes highly counter-intuitive wave phenomena observed in the Sun.

Accurately deconstructing the sub-surface structure and dynamics of active regions is a difficult task. Since the development of methods of time-distance helioseismology \citep{duvall,gizon05} and the subsequent investigations into the nature of the sunspot underbelly \citep{duvall96,kosovichev,couvidat}, there have been several arguments attempting either to establish the significance of MHD interactions in sunspot structure and dynamics inversions \citep[e.g.][]{lindsey,schunker} or to the contrary \citep{zhao}. Recent theories \citep{aaron} argue that most of the observed wave phase shifts in sunspot regions occur in a thin sub-photospheric region of 1 Mm depth, where magnetic field effects are putatively the largest. The implication is that the causative mechanisms behind observed wave phase shifts may have been misidentified, a conclusion echoed by \citet{hanasoge4} who demonstrate that wave source suppression due to convective blocking in sunspots can also participate in the task of creating time shifts \citep[also see,][]{gizon02}. Moreover, wave phase shifts inferred in regions of strong magnetic fields from Michelson Doppler Imager \citep[MDI;][]{scherrer} observations \citep[e.g.][]{duvall96} are difficult to interpret because of substantial changes in the line formation height due to profound alterations in the thermal structure of the underlying plasma. On the positive side, the prevalence of computing resources and numerical methodology now afford us the ability to conduct investigations that may not have been possible a decade ago. Developing an interaction theory of waves and magnetic fields will allow more consistent studies of sunspot structure and dynamics. 

The reduction in acoustic oscillation power in sunspot regions \citep[e.g.][]{lites} has been the subject of extensive observations with several theories put forth to explain this phenomenon \citep[e.g.][]{hindman97,parchevsky007}. \citet{hindman97} have discussed several plausible mechanisms that may be contributing to the power reduction but the participatory extents are as yet unknown. On a related issue, a number of studies have focused on placing observational constraints on the degree of wave absorption in sunspots \citep[e.g.][]{braun87,bogdan93,braun95,cally95}. The technique discussed here provides an independent manner of investigating all these issues. Acoustic or seismic enhancements (or halos as they are termed in this paper) are ubiquitously seen in both velocity and intensity observations, encircling active regions \citep[e.g.][]{braun92,brown92,balthasar,hindman98,donea,nagashima}. Some \citep{brown92,donea} have speculated that they originate from enhanced source activity in the vicinity of the active region. In this paper, we present power maps from simulations of waves interacting with moderate to strong magnetic fields; acoustic halos are clearly seen in these images, implicating an MHD based mechanism. 

On a very different scale but of equal importance are small magnetic elements and thin flux tubes. The dynamical emergence and disappearance of these flux tubes provides us insights into the photospheric dynamo \citep[e.g.][]{cattaneo}. In a bid to understand the structure of these flux tubes, \citet{duvall06} analyzed MDI observations of thousands of independent small magnetic elements, thereby developing a highly resolved statistical picture of the associated wave scattering. Understanding the nature of the interaction between thin flux tubes and waves may allow us to recover details of the flux tube structure from the scattering information. Forward models of wave interactions with thin flux tubes \citep[e.g.][]{bogdan95,bogdan96,gizon06,hanasoge5} can then be constructed in order to place restrictions on the subsurface magnetic field distribution. Models of this sort can be used in theoretical studies of flux emergence \citep[e.g.][]{cheung}.

In this regard, a first step is to devise a sufficiently general manner of computing wave propagation in a magnetized plasma. The linearized ideal MHD equations provide a reasonable starting point, since MHD oscillations in the photosphere and below are governed by predominantly linear physics \citep[e.g.][]{bogdan}. \citet{cally97}, \citet{rosenthal}, and  \citet{cally00} performed MHD simulations in two dimensions to study rates of mode absorption in magnetic flux tubes. Subsequently, \citet{cameron07} developed and validated numerical techniques to perform 3D linear MHD computations with a focus on recovering the magnetic field distribution based on wave scattering measurements. The assumption of linear wave propagation and time stationarity of the background state are common threads between this work and that of the above-cited authors.

High-order numerical accuracy is a minimum requirement for computational work. The linear calculation discussed here does not face the same restrictions as would a non-linear counterpart, where the presence of shocks makes it quite difficult to raise the order of the numerical scheme without introducing instabilities. We discuss the methods employed to spatio-temporally evolve solutions of the ideal MHD equations in $\S$\ref{comp.method.sec}. Subsequently, an empirical method to generate stable MHS states is introduced in $\S$\ref{mhs.states.sec}, with an illustration of one such state: a flux tube with peak photospheric field strength 600 G. Results of wave simulations with some flux tubes, specifically the phenomena of wave power reduction and enhancement are discussed in $\S$\ref{results.sec}. Finally, we summarize and conclude in $\S$\ref{conclude.sec}.

\section{COMPUTATIONAL METHOD\label{comp.method.sec}}
\begin{figure}[!ht]
\begin{centering}
\epsscale{0.75}
\plotone{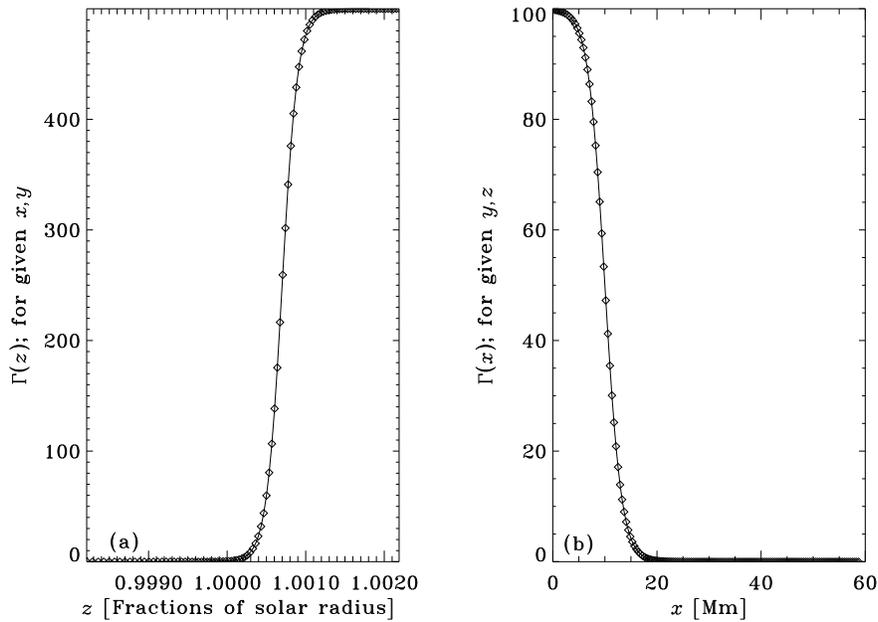}
\caption{The function $\Gamma(x,y,z)$ of equations~(\ref{cont}) - (\ref{induction}). In panel a, the value of $\Gamma$ at a location far away from the side boundaries is plotted as a function of $z$. The vertical boundaries of the computational box are at $z=0.95, 1.002 R_\odot$; although not shown here, the behavior of $\Gamma$ at the lower boundary is qualitatively similar to that in panel a. In panel b, the variation of $\Gamma$ with the horizontal coordinate $x$ at a location far away from the vertical and $y$ boundaries, with $x=0$ serving as one of the side boundaries. As waves approach within 20 Mm of the horizontal and/or 1 Mm of the vertical boundaries, they start to experience strong damping by the $-\Gamma$ term.\label{Gammafig}}
\end{centering}
\end{figure}

Similar to the forward models of the solar wave field developed in \citet{hanasoge1} and \citet{hanasoge3}, we start by linearizing and modifying the ideal MHD equations in the following manner:
\begin{equation}
\partial_t\rho = -\bnabla\dotp(\rho_0\vel) -\Gamma\rho,\label{cont}
\end{equation}

\begin{equation}
\partial_t\vel = -\frac{1}{\rho_0}\bnabla p - \frac{\rho}{\rho_0} g \zhat + \frac{\zeta(z)}{4\pi\rho_0}[(\bnabla\curl\bB_0)\curl\bB + (\bnabla\curl\bB)\curl\bB_0] + {\bf S}-\Gamma\vel,\label{mom}
\end{equation}

\begin{equation}
\partial_t p =  - \vel\cdot\bnabla p_0 -\rho_0 c^2\bnabla\dotp\vel-\Gamma p,\label{energy}
\end{equation}

\begin{equation}
\partial_t\bB = \zeta\bnabla\curl\left(\vel\curl\bB_0\right) -\Gamma\bB \label{induction}
\end{equation}

\begin{equation}
\bnabla\cdot\bB = 0\label{delb}
\end{equation}
where $\rho$ denotes density (unless stated otherwise, the subscript `0' indicates a time-stationary background quantity while un-subscripted terms fluctuate), $p$ pressure, $\bB = (B_x,B_y,B_z)$ the magnetic field,  $\vel=(v_x,v_y,v_z)$ is the vector velocity, $g =g(z)$ is gravity with direction vector $-\zhat$, $c = c(x,y,z)$ is the sound speed, $\Gamma=\Gamma(x,y,z) > 0$ is a damping sponge that enhances wave absorption at all horizontal and vertical boundaries (see Figure~\ref{Gammafig}), $\zeta(z)$ a Lorentz force `controller' (Robert Cameron, private communication 2007; Robert Stein, private communication 2007), and ${\bf S}$ is the source term. The controller term $\zeta$ (see Figure~\ref{zeta.dep}a) is such that it is constant (=1) over most of the interior but decays rapidly with height above the photosphere. Note that $\zeta$ is also present in equation~(\ref{induction}) - as the influence of the magnetic field on the fluid decreases (Eq.~[\ref{mom}]), so must the effect of the fluid on the magnetic field. For further discussion on the reasoning behind this term, see $\S$\ref{controller}.

We employ a Cartesian coordinate system $(x,y,z)$ with $\zhat$ denoting the unit vector along the vertical or $z$ axis and $t$, time. Because of the presence of a spatially varying magnetic structure, the background pressure, density, and sound speed adopt a full three-dimensional spatial dependence. In sequential order, equations~(\ref{cont}) through~(\ref{energy}) enforce mass, momentum, and energy conservation respectively, while equation~(\ref{induction}) is the induction equation. Equation~(\ref{delb}) assures us that magnetic monopoles do not exist. In interior regions of the computational box (away from the boundaries), solutions to the above equations are adiabatic since the damping terms decay to zero here. The source term ${\bf S}$, is a spatio-temporally varying function, the structure of which has been discussed in some detail in \citet{hanasoge2} and \citet{hanasoge3}. Essentially, it is a phenomenological model for the multiple source wave excitation picture that is observed (inferred perhaps) in the Sun. The background vertical stratification is an empirically derived \citep{hanasoge1}, convectively stabilized form of model S \citep{jcd}. 

The base hydrodynamic method remains unchanged from \citet{hanasoge3}; spatial derivatives are calculated using sixth-order compact finite differences \citep{lele} and time evolution is achieved through the repeated application of an optimized second-order five-stage Runge-Kutta scheme \citep{berland}. The temporal order of accuracy is dropped because the time step (2 seconds) is much smaller than the period of the waves studied here. The boundaries are lined with damping sponges in order to absorb (damp) outgoing waves (Figure~\ref{Gammafig}). This is to prevent any scattered waves from re-entering the computational domain as would be the case with periodic boundaries. Our attempt to extend the base scheme to compute the magnetic field terms in equations~(\ref{mom}) and~(\ref{induction}) was successful. All derivatives, including the magnetic field terms, are estimated using sixth-order compact finite differences, thus maintaining a high order of spatial accuracy. It was observed that the $\bnabla\cdot{\bf B}$ term was of a low magnitude, $\lesssim 10^{-7}$ per pixel, and therefore harmless \citep[e.g.][]{toth,abbett}. Moreover, the presence of the damping term $\Gamma \bB$ ensures that $\bnabla\cdot{\bf B}$ is forced to decay in the damping sponge layers. Validation in one and two dimensions of the essential numerical method (i.e. without the $\zeta$ or $\Gamma$ terms) is discussed in Appendix~\ref{validation.sec}.

\subsection{Lorentz force controller}\label{controller}
\begin{figure}[!ht]
\centering
\epsscale{1.0}
\plotone{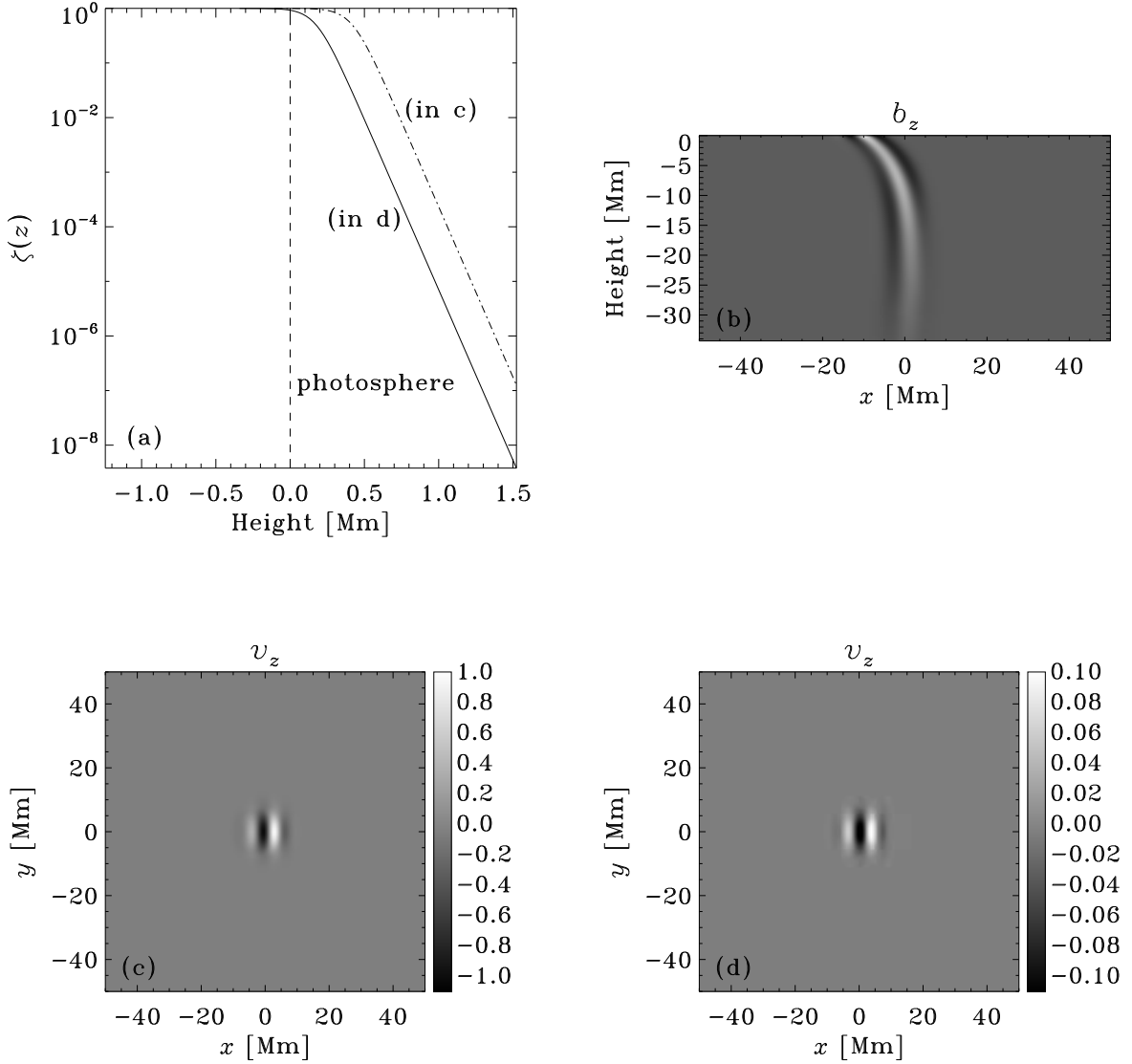}
\caption{Figure showing the dependence of wave scattering on $\zeta(z)$ (Eq.~[\ref{mom}],~[\ref{induction}]). Panel a shows functional forms of $\zeta$, used in two different MHD wave test simulations, termed c and d. The initial condition, same for both simulations, is a plane wave packet localized at $x=-30$ Mm, $z=-0.2$ Mm. Panel b shows fluctuations in $b_z$, which arise due to the interaction of the wave with the flux tube. Panels c and d display the instantaneous normalized vertical velocity ($v_z(x,y,z=0.2~{\rm Mm},t=31~ {\rm min})$, units are arbitrary) of the scattered waves extracted at a height of 200 km above the photosphere from simulations c and d respectively. An order of magnitude difference is seen between the two cases, indicating that the results are somewhat sensitively dependent on the chosen form of $\zeta$.\label{zeta.dep}}
\end{figure}

As stated in $\S$\ref{comp.method.sec}, $\zeta$ retains the value 1 in the interior and decreases with height above the photosphere (Figure~\ref{zeta.dep}a). It attempts to achieve a two-fold purpose: (I) a reduction in the Lorentz force with increasing altitude above the photosphere and (II) prevent the onset of negative pressure effects. The mean hydrodynamic pressure and density in Sun drop exponentially with height in the atmosphere that immediately overlays the photosphere. In calculations of MHS states ($\S$ \ref{mhs.states.sec}), it was nearly impossible to prevent complete pressure and density evacuation in the interiors of flux tubes of large magnitude field strengths \citep[1500 Gauss and more - sadly nowhere close to the umbral field strengths of up to 6100 G that have been observed in sunspots by][]{living}. Moreover, the equilibrium horizontal pressure distribution takes on strange forms, with the pressure at the center of flux tube attaining larger values than the ambient, when the flux tube radius is forced to increase faster than the corresponding potential field configuration. In the Sun, the presence of magnetic field everywhere and the phenomenon of flux tube merging in the atmosphere \citep[e.g.][]{pneuman,bogdan96} help reduce large gradients in the magnetic field, thereby preventing complete evacuation in active regions and sunspots while not requiring the flux tubes to flare out too rapidly. We attempt to simulate this (criterion I) through the $\zeta(z)$ term. However, since the equilibrium structure of an active region is as yet unknown, it is not possible to determine how realistic a chosen functional form of $\zeta$ is. 

To determine the impact of $\zeta(z)$ on the wave field, we simulate the interaction of a wave packet with a relatively weak flux tube ($\sim$ 100 Gauss at the photospheric level) in a solar-like stratified medium. Three simulations are performed, a quiet simulation (`q') without any magnetic field and two (`c' and `d') with different functional forms of $\zeta$, one form of $\zeta$ decaying more rapidly with height than the other (shown in panel a of Figure~\ref{zeta.dep}). The initial condition for all simulations was chosen to be a Gabor wavelet shaped disturbance in $v_z$ localized at $(x,z)= (-30,-0.2)$ Mm, at all $y$. At approximately the instant when the wave packet reaches the center of the flux tube (located at $(x,y) = (0,0)$ Mm), we display snapshots of $v_z^{\rm c}-v_z^{\rm q}$ and $v_z^{\rm d}-v_z^{\rm q}$ in panels c and d respectively, where the superscripts refer to the simulation index (q,c, or d). In the presence of a linear scatterer, one may view the velocity field as being associated with an incident and scattered wave; in this situation, $v_z^{\rm q}$ is the incident wave velocity, while the scattered wave velocities are described by the differences $v_z^{\rm c,d}-v_z^{\rm q}$. It is clear from panels c and d of Figure~\ref{zeta.dep} that the extent of scatter in the simulation where $\zeta$ decays higher up in the atmosphere (c) is greater by an order of magnitude than (d). Perhaps mode conversion, which has been theoretically shown to become significant when the plasma-$\beta$ starts to drop, is at play \citep[e.g.][]{bogdan96,cally97,crouch03}. It may also be that the magnetic field changes the $v_z$ eigenfunction more significantly in one case than the other. Essentially, this experiment tells us that capturing wave interactions in an active regions is somewhat sensitively dependent on the choice of the $\zeta$ function, or in other words, on the atmospheric magnetic field distribution in the vicinity. It underlines the neccessity of viewing this effort as more qualitative than quantitative, since conclusions of the latter sort require exploring a formidable parameter space.

\section{MHS STATES\label{mhs.states.sec}}
\begin{figure}[!ht]
\centering
\epsscale{1.0}
\plotone{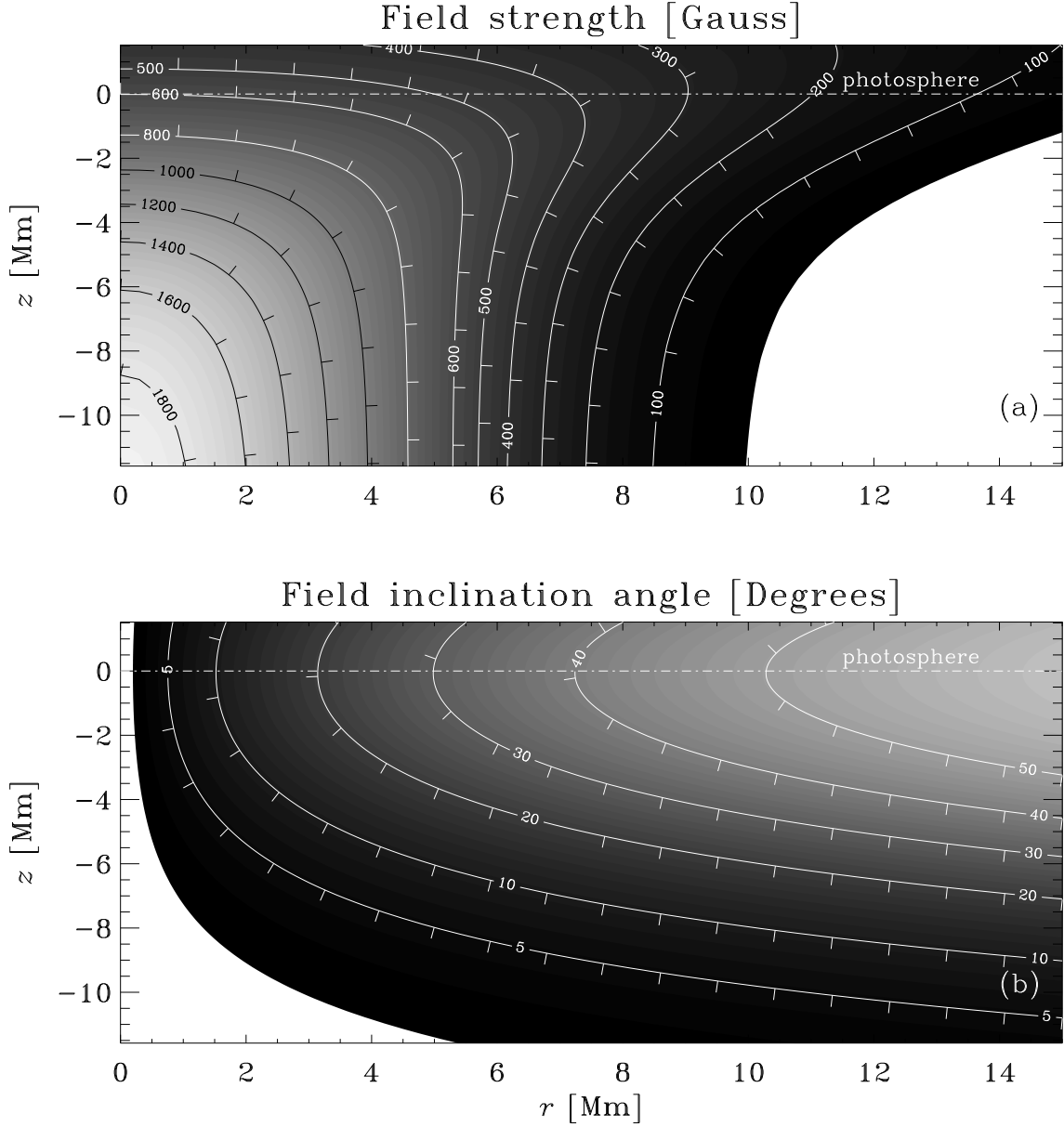}
\caption{An example of a flux tube generated according to the recipe of $\S$\ref{mhs.states.sec}. Panel a shows the field strength, $|B| = [B_r^2 + B_z^2]^{1/2}$; panel b shows the field inclination atan$(B_r/B_z)$. Perpendicular to the contour lines are spokes that point in the direction of the downhill gradient. The pressure and density remain positive over the entire domain. Field strength magnitudes in Gauss and inclination angles in degrees are indicated along the contour lines. \label{fluxtube.examp}}
\end{figure}

Generating MHS states in stratified media can be a non-trivial task \citep[e.g.][]{pneuman,pizzo,belien,khomenko}. Fully consistent approaches that involve relaxing the MHD equations to low-energy equilibria are difficult to implement. Moreover, such calculations are beyond the scope of this effort; we are interested less in the MHS state itself than in the manner in which waves interact with them. We invoke the \citet{schluter} self-similar magnetic field geometry and ignore both radiative transfer effects and the satisfaction of the equation of state. We also remind the reader that the background stratification has been altered to prevent the onset of uncontrolled linear growth of convective instabilities, thus changing the opacities in a non-physical manner. The \citet{schluter} approximation tells us that making the following choices for the radial and vertical magnetic field $B_r$ and $B_z$ assures us of the satisfaction of equation~(\ref{delb}) \citep[e.g.][]{schussler}:
\begin{equation}
B_z = M \psi(z) e^{-r^2 \psi(z)},\label{eq.bz}
\end{equation}
\begin{equation}
B_r = - M\frac{r}{2} \psi' e^{-r^2 \psi(z)},\label{eq.br}
\end{equation}
with $\psi' = d\psi/dz$. The above equations~(\ref{eq.bz}) and~(\ref{eq.br}) are in cylindrical geometry; $r$, $z$ refer to the horizontal radial and vertical coordinates with $r=0$ coinciding to the center of the flux tube, $M$ a term that controls the magnitude of the magnetic field and hence the flux ($=\pi M$), and $\psi(z)$, the horizontal extent of the flux tube and the rate at which the flux tube spreads with altitude. The zeroth-order MHS equations in cylindrical coordinates, obtained upon dropping the time and azimuthal dependencies in equation~(\ref{mom}), reduce to:
\begin{equation}
0 = -\partial_r p + \zeta\frac{B_z}{4\pi} \left[ \partial_z B_r - \partial_r B_z\right] \label{eq.pr},
\end{equation}
along the horizontal ($r$) direction and in the vertical ($z$) direction,
\begin{equation}
0 = -\partial_z p - \zeta\frac{B_r}{4\pi} \left[ \partial_z B_r - \partial_r B_z\right] - \rho g.\label{eq.pz}
\end{equation}
Equation~(\ref{eq.pr}) is integrated from $r = 0, \infty$ to obtain the following equation:
\begin{equation}
p_c(z) = p_\infty(z) + \frac{M^2\zeta}{4\pi}\left[\frac{1}{16}\frac{\psi'^2}{\psi} - \frac{1}{8}\psi'' + \frac{\psi^2}{2}\right],\label{eqb.temp}
\end{equation}
where $p_c(z)$ is the pressure along the axis (centerline) of the flux tube and $p_\infty(z)$ is the hydrostatic pressure far away from the magnetic region. The horizontal pressure distribution at a given $z$ can now be computed by integrating equation~(\ref{eq.pr}) from the center outward:
\begin{equation}
p(r',z) = p_c(z) + \frac{\zeta}{4\pi}\int_0^{r'} dr B_z  \left[ \partial_z B_r - \partial_r B_z\right]\label{eqb.pr};
\end{equation}
thus the entire pressure field can be recovered through this procedure. Simplifying equation~(\ref{eq.pz}), we can obtain the density field from the pressure distribution:
\begin{equation}
\rho(r,z) = -\frac{1}{g}\left( \partial_z p + \zeta\frac{B_r}{4\pi} \left[ \partial_z B_r - \partial_r B_z\right]\right).\label{eqb.den}
\end{equation}
Therefore, upon specifying parameters $M$ and $\psi(z)$ in equations~(\ref{eq.bz}) and~(\ref{eq.br}), one can obtain a self-consistent MHS solution that satisfies the criteria of $\bnabla\cdot\bB = 0$ and magneto-hydro-static balance. One must be careful however to ensure positive pressure and density in equations~(\ref{eqb.pr}) and~(\ref{eqb.den}) at all points in the computational domain. In Figure~\ref{fluxtube.examp}, we show an example of a flux tube which attains a peak strength of 600 G at the photospheric level; the inclination of the field at distances away from the center is also shown.

\begin{figure}[!ht]
\centering
\epsscale{1.0}
\vspace{-6cm}
\plotone{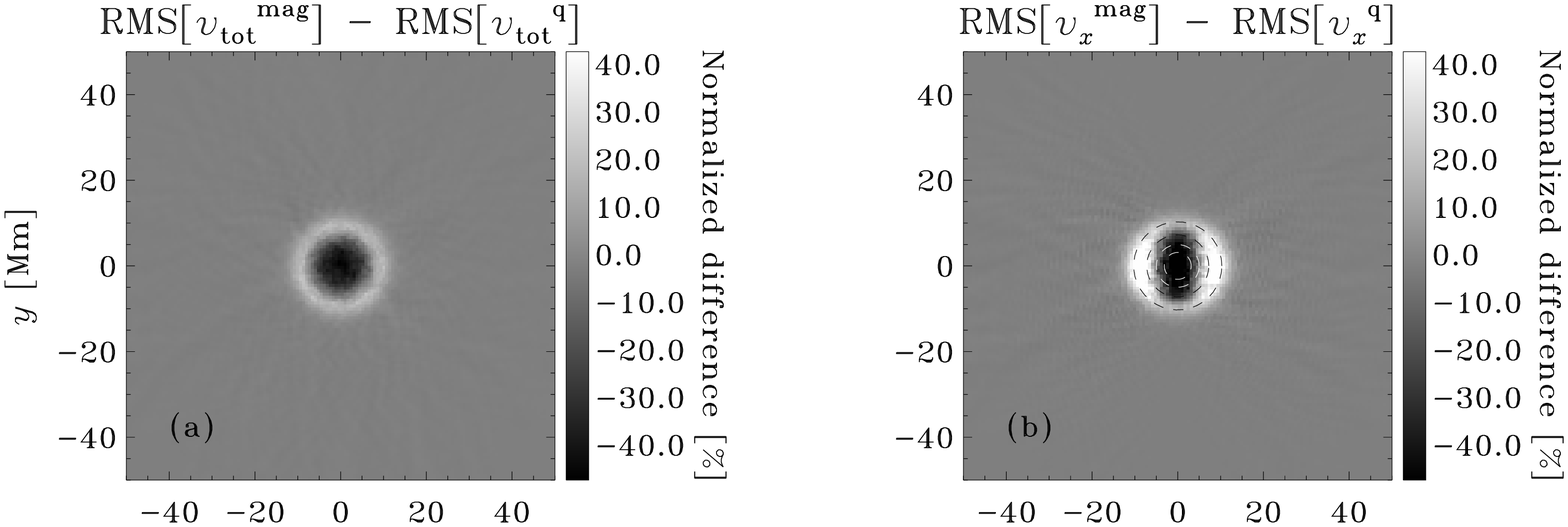}
\vspace{-11cm}
\plotone{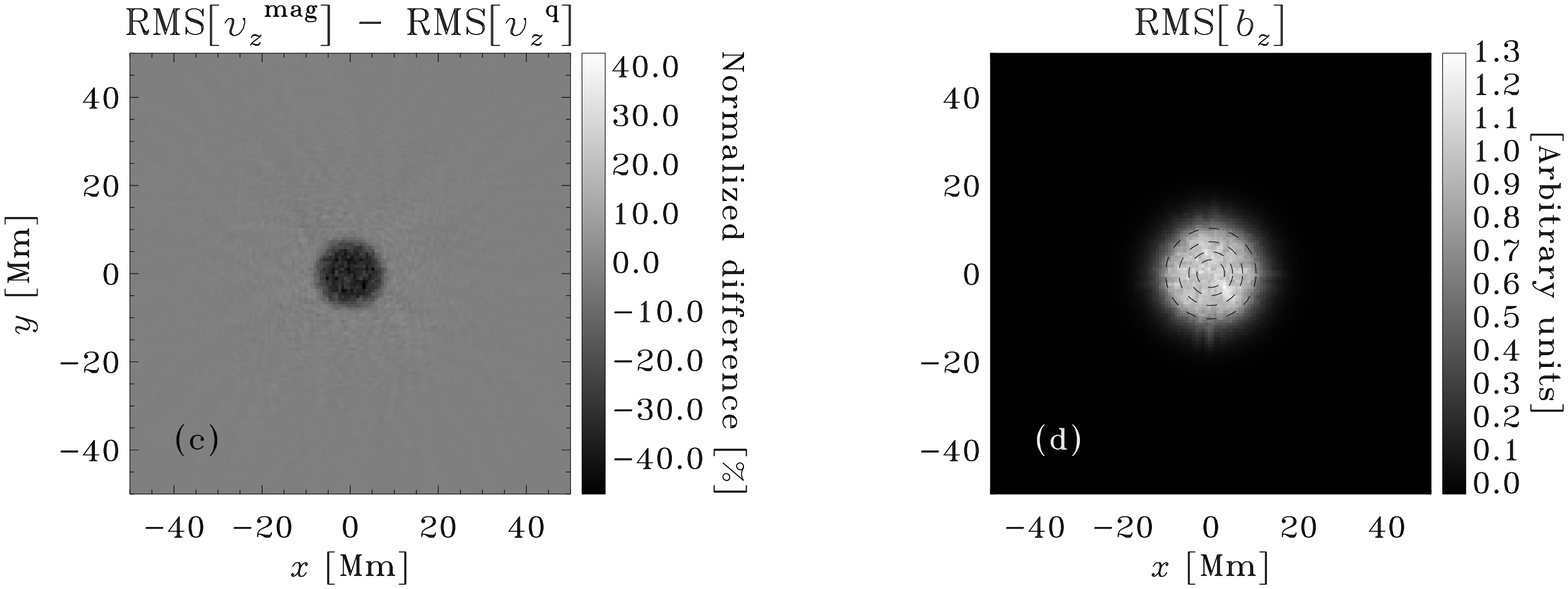}
\vspace{-6cm}
\caption{Changes in acoustic power caused due to MHD interactions (600G case). Two simulations were performed: a quiet (`q') and an MHD calculation (`mag') with the flux tube of Figure~\ref{fluxtube.examp} embedded in the computational domain. Panel a shows the RMS differences of the total velocity, $v_{\rm tot} = \sqrt{v^2_x + v^2_y + v_z^2}$ between the quiet and magnetic simulations. Panels b and c display the RMS differences seen in $v_x$ and $v_z$ while d is the RMS of the vertical magnetic field fluctuations, $b_z$. Each power difference is normalized by the mean value of the RMS of the corresponding quantity (i.e. $v_x, v_z$, or $v_{\rm tot}$) derived from the quiet simulation. With increasing radius, the contours in panels b and d show locations where the field inclination is $[20^\circ, 30^\circ, 40^\circ, 50^\circ]$. The field strengths at these contours are [560, 495, 397, 240] G respectively. Outside a region of substantial decrease in wave oscillation amplitudes, a halo corresponding to an increase in the RMS is seen.\label{rms.changes}}
\end{figure}

\subsection{Seismic Power Deficits and Halos}\label{results.sec}
Theoretical expectations dictate a decrease in modal power in magnetic regions due to mode absorption and MHD-wave coupling. Using identical realizations of the source function ${\bf S}$ (Eq.~[\ref{mom}]), we perform two simulations: a `quiet' run and an MHD counterpart (`mag') with the flux tube of Figure~\ref{fluxtube.examp} embedded at the center of a computational box of size $100 \times 100 \times 35~{\rm Mm}^3$ via the computational method of $\S$\ref{comp.method.sec}. In Figure~\ref{rms.changes}, we show the difference in time-averaged Root Mean Square (RMS) wave power between a quiet simulation and its magnetic counterpart, normalized by the mean value of the RMS of the quiet case. Both runs were twelve hours long. Because the realizations are identical, the MHD interactions are the dominant component of the quantity RMS$^{\rm mag}$ - RMS$^{\rm q}$. It is interesting to note that depending on the variable of study, the simulations predict strong variations in the nature and degree of change in wave power. For example, the RMS differences in the total velocity, $v_{\rm tot} = \sqrt{v_x^2 + v_y^2 + v_z^2}$ show the presence of a large reduction in wave power surrounded by an intense halo, whereas the RMS decrease as seen in $v_z$ is systematically weaker and an almost invisible halo. The panels b and d show contours of increasing radii corresponding to field inclinations of $[20^\circ, 30^\circ, 40^\circ, 50^\circ]$. The halo is seen at inclinations of $50^\circ$ and higher, while a strong reduction in wave power is observed at smaller angles. Also, the robustness of the halo was ensured by verifying its reappearance in a simulation using an alternate numerical method, namely a second-order Constrained-Transport technique \citep[CT;][]{evans}.

\begin{figure}[!ht]
\centering
\epsscale{1.0}
\plotone{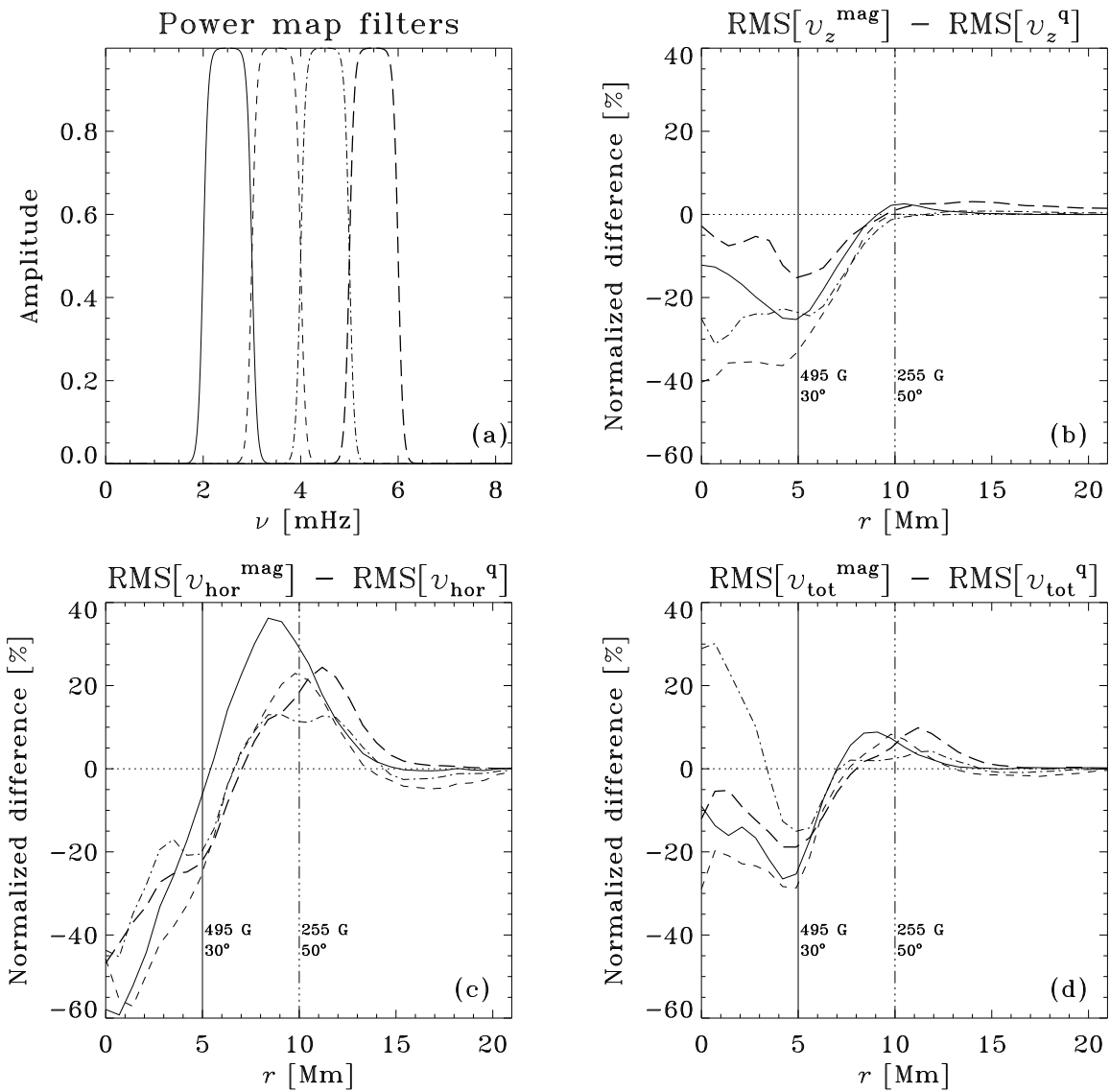}
\caption{Azimuthal averages of the frequency filtered wave power maps (600G case). Panel a shows four filters with bandpasses 2 - 3, 3 - 4, 4 - 5, and 5 - 6 mHz. Panels b, c, and d display the azimuthally averaged (about the center of the flux tube) normalized noise-subtracted power maps of the quantities $v_z, v_{\rm tot}$, and $v_{\rm hor} = \sqrt{v_x^2 + v_y^2}$. Each power difference is normalized by the mean value of the RMS of the correspondingly filtered quiet simulation. $v_z$ exhibits the least change in the RMS of all the variables shown here. Note that $v_{\rm hor}$ and $v_{\rm tot}$ are more difficult to interpret because sign information is lost ($v_{\rm hor}^{\rm q, mag},v_{\rm tot}^{\rm q, mag} \ge 0$). The two lines parallel to the $y$-axis in panels b,c, and d show the magnetic field strength and inclination at these locations.\label{azi_ave}}
\end{figure}

In order to study these effects further, we computed power maps in four different frequency bandpasses, 2 - 3, 3 - 4, 4 - 5, and 5 - 6 mHz. Different components of the velocity were used in the calculations, $v_z, v_{\rm tot},$ and $v_{\rm hor} = \sqrt{v_x^2 + v_y^2}$. We subtract the power maps of the quiet simulation computed in the same bandpasses to reduce the realization noise. The frequency filters used to recover the power maps and the azimuthally averaged power profiles (about the flux tube center) obtained subsequently are shown in Figure~\ref{azi_ave}. Noteworthy aspects are that $v_{\rm hor}$ contains the most intense halos, $v_{\rm tot}$ shows a dramatic increase in RMS power in the range 4 - 5 mHz around the `umbral' region of the flux tube (defined as within a distance of 8 Mm from the center of the flux tube), while $v_z$ displays limited shifts in the RMS in comparison to the rest. These effects (or some fraction thereof) could be attributed to changes in the eigenfunctions caused by the magnetic fields. The appropriate identification of the nature of these increments and decrements is evidently an important issue.

\begin{figure}[!ht]
\centering
\epsscale{1.0}
\vspace{1cm}
\plotone{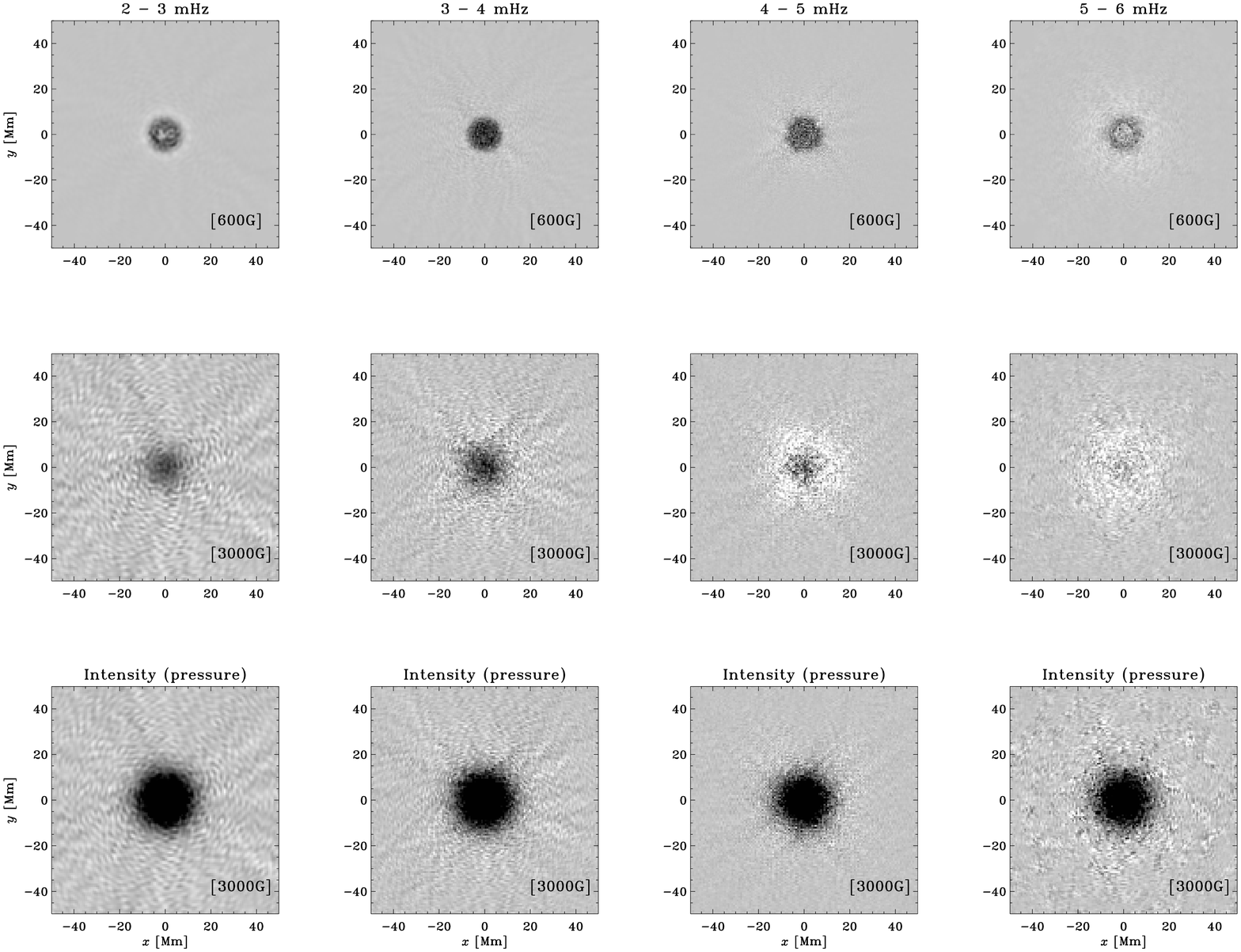}
\caption{Changes in acoustic power caused due to MHD interactions as observed in $v_z$, the vertical velocity (upper two rows), and $p$, the pressure. The color scale is fixed to the range $[-50, +15]$\%, where each map has been normalized by the value of the quiet power in that frequency range. In both simulations, the source distributions are spatially uniform. The qualitative and quantitative differences seen between the two cases (images in the upper two rows) could be ascribed to the magnetic field strength, the location of $\beta=1$ layer, and the source depth. We use pressure fluctuations as a proxy for observations in intensity. Although more careful studies are required to deconstruct these results into participatory elements, it can be said that the computations reproduce many features in the velocity observations \citep[e.g.][]{moretti,nagashima}.\label{fat_tube}}
\end{figure}

Another set of power maps is displayed in Figure~\ref{fat_tube}. The upper set of panels contains the power maps of the 600 G flux tube whereas the lower two rows show the results from a simulation with a more realistically endowed sunspot: a 3000 G flux tube (simulation size: $200\times200\times35~{\rm Mm}^3$). The flux tube configuration is very similar to that discussed in \citet{cameron07}; consequently, we do not show it here. The middle images are strikingly similar in structure to the observations of \citet{moretti}, who see power increasing progressively with frequency (Figure 1 of their paper). We see a large decrease in the RMS power as felt by the pressure fluctuations (interpreted crudely as intensity) in the lowest set of panels. There is some qualitative agreement between the simulations and the intensity observations by \citet{moretti}; however, the high resolution Hinode measurements of intensity in active regions by \citet{nagashima} are unfortunately not so easily woven into this computational web. Intensity observations, as note \citet{nagashima}, are far more difficult to interpret than those in velocity because of its sensitivity to the ionization, pressure, density etc., and the lack of a one to one correspondence with a simple thermodynamic variable.

Acoustic halos around the edges of active regions have been widely observed \citep[e.g.][]{braun92,brown92,balthasar,donea}. While \citet{balthasar} have reported enhancements in oscillation velocity power also within magnetic regions and in low frequency ($\sim 2$ mHz) bandpasses, a large number of other observations seem to show halos only in a high frequency bandpass and in predominantly weakly magnetic areas surrounding the active region \citep{hindman98}. It is interesting to note that some qualitative features also seen in observations are reproduced in the simulations: (1) at the edge of the flux tube (at $\sim 19 - 20$ Mm in Figure~\ref{azi_ave}, $|\bB| \sim 7 - 12$ G), only the highest frequency bandpass shows a faint power enhancement, of the order of 2 - 3 \% in $v_z$ and even less in the other components, (2) the increase in wave power in the umbra of the flux tube in the 4 - 5 mHz bandpass (panel d of Figure~\ref{azi_ave}) is similar to enhancements seen in the magnetic cores of active regions \citep{balthasar}, and (3) the enhancements grow with frequency, as seen in the simulation of the 3000 G flux tube of Figure~\ref{fat_tube} and in observations by \citet{moretti}. It would be rather ludicrous to make quantitative comparisons between observations and the simulations because of the simplified nature of these calculations: the lack of radiative heat transfer, realistic wave mode damping, a penumbra, convection, unmodeled atmospheric magnetic fields etc. 

The speculation that enhanced seismic emission in the vicinity of active regions may be the causative mechanism of the acoustic halo goes back to the work of \citet{brown92}. More recently, \citet{donea} have drawn similar conclusions from holography related analyses of active region observations. However, this theory does not explain the wave power increase in the simulations because in our calculations, wave source amplitudes are statistically homogeneously distributed in space (in the horizontal directions) with the exception of areas close to the boundaries. Magnetic regions reconfigure the energy of the background medium. Therefore, the presence of sources in the interior of the flux tube essentially complicates matters because the incipient waves may have energies unlike waves in quiet regions. Moreover, the relative locations of the $\beta =1$ layer with respect to the acoustic reflection zone, the $\tau=1$ line, and the sources probably play an extremely important role in determining the wave energy distribution as a function of frequency. All the variables in the simulation are extracted at constant geometrical height (200 km above the photosphere), clearly a simplification incongruent with reality. Whether the observation height is a significant contributor is yet to be determined. Further investigations are currently in progress and will be the focus of a future paper.

\section{DISCUSSION\label{conclude.sec}}
We have discussed and validated a numerical method to systematically study linear MHD interactions in the context of helioseismology. The importance of including the ambient atmospheric magnetic field in the vicinity of magnetic flux concentrations is underlined here. Through a phenomenological model of the gradient smoothing that the ambient magnetic field presumably effects, we have shown that there can be significant differences in estimates of the oscillation velocity inside active regions. Thus, forward models that attempt to recover the magnetic field distribution based on shifts in travel times or other helioseismic metrics must in fact address this issue. Computational studies pertaining to oscillation power reduction in active regions are also quite sensitive to these effects.

Results from simulations of waves interacting with 600 and 3000 G strong flux tubes are discussed in some detail. Not only is a significant reduction in wave power observed but a halo that surrounds the flux tube is also seen. Many features in the velocity observations of active regions are reproduced by the simulations. High frequency wave power halos are also observed to envelope solar active regions; \citet{braun92}, \citet{brown92} and \citet{donea} suggest enhanced seismic emission in the vicinity as being the causative mechanism. However, the simulations contain no such seismic enhancements, indicating that the physics behind the formation of the halo is possibly governed by MHD phenomena. A theory to explain the appearance of these excess oscillations will be discussed in a future publication. 

Using the techniques described here, we wish to develop helioseismically consistent forward models of thin flux tubes and sunspots. In the context of thin flux tube models, preliminary investigations have already shown that the peak flux tube magnetic field strengths of about 80 G \citep{duvall06} as observed by the MDI instrument are too small by two orders of magnitude to cause the observed wave phase shifts. This is a consequence of the relatively low resolution of MDI, which is unable to capture the 100-200 km sized flux tubes (Tom Duvall, Jr. 2007; Tom Bogdan 2007; Robert Cameron 2007, various private communications). Simulations with such small features can be computationally challenging due to resolution restrictions and the associated computational overhead. However, interesting sub-wavelength physics associated with thin flux tubes, namely the near-field evanescent modes \citep[the $jacket$, e.g.][]{bogdan95,hanasoge5} can be studied in greater detail with these simulations. These investigations are exciting, especially seen in the context of the availability of high quality observations and the upcoming Solar Dynamics Observatory (SDO) mission.

\appendix
\section{Validation: 2D analytical solution}\label{validation.sec}
\begin{figure}[!ht]
\centering
\epsscale{0.75}
\plotone{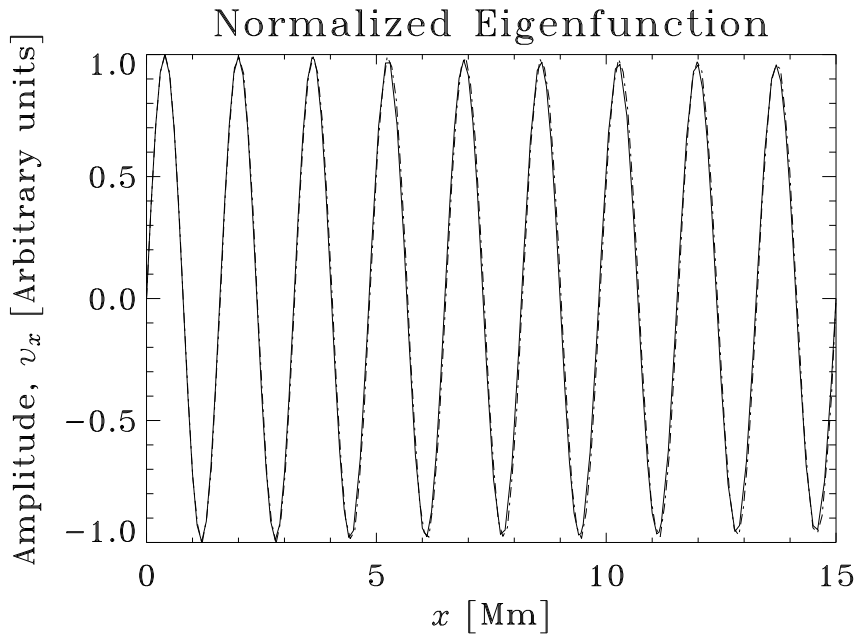}
\caption{The analytically computed (solid line) and numerically simulated (dot-dash line) eigenfunctions for $\nu=5.09$ mHz, $k=0$. For convenience, the background magnetic was chosen as $B_0 = \tilde{b} \sqrt{2x}$. Although not shown here, we have also tested the simulation at non-zero values of $k$ and found good agreement.\label{eigfunc.examp}}
\end{figure}

Take a 2D slab of finite thickness $(L,L)$. Let the coordinates be labeled $(x,z)$ and assume the presence a background magnetic field of the form ${\bf B_0} = \bozx\ez$. The background density is assumed to be unchanged by the magnetic field and is spatially non-varying; the pressure $p_0$ is adjusted so that a pressure balance is achieved. We choose a velocity of the form, ${\bf v} = (v_x\ex + v_z\ez)\exp\left[i(kz - \omega t)\right]$, where $v_x = v_x(x), v_z= v_z(x)$, $k$ the wavenumber, $\omega$ the frequency, and $t$, time. Background quantities are denoted by the subscript 0. The magnetic field and pressure fluctuations are denoted by ${\bf B}$ and $p$ respectively. Since this solution is used to validate the code, we use the linearized ideal MHD equations, which are equations~(\ref{cont}) through~(\ref{delb}) without the boundary dissipative $\Gamma$ or Lorentz force controller $\zeta$ terms; we also set the source term ${\bf S} = 0$. Starting with the adiabatic energy equation (upon incorporating the continuity equation), we have:
\begin{eqnarray}
\partial_tp &=& -c^2\rho_0\bnabla\dotp{\bf v} - v_x\partial_x p_0, \\
c^2 &=& \frac{\Gamma_1 p_0}{\rho_0} \\
\partial_x p_0 &=& -\db, \label{pressure.balance}\\
-\frac{1}{\rho_0}\partial_xp &=& \frac{1}{i\omega\rho_0}\left[\Gamma_1\db(\partial_xv_x + ikv_z) - \Gamma_1 p_0(\partial^2_xv_x + ik\partial_xv_z)  \right.\nonumber\\ &+& \left. \partial_xv_x\db + v_x\ddb\right]\ekx,\label{pressure.x}\\
-\frac{1}{\rho_0}\partial_z p &=& \frac{k}{\omega\rho_0} \left[-\Gamma_1 p_0 (\partial_x v_x + ik v_z) + v_x\db\right]\ekx\label{pressure.z},
\end{eqnarray}
where equation~(\ref{pressure.balance}) is the pressure distribution created by balancing the Lorentz force due to the background magnetic field and $\Gamma_1$ is the first adiabatic index. Moving on to the $x$-momentum equation, and applying equation~(\ref{pressure.x}), 
\begin{eqnarray}
v_x &=& \frac{1}{\rho_0 \omega^2} \left[ \Gamma_1\db(\partial_xv_x + ikv_z) - \Gamma_1 p_0 (\partial^2_x v_x + ik \partial_xv_z) + \partial_x v_x\db\right. \nonumber\\ &+& \left. v_x\ddb + k^2 v_x B^2_0 - B_0\partial^2_x(v_x B_0) -\partial_xB_0\partial_x(v_xB_0)\right]. \label{eq.vvx}
\end{eqnarray}
Similarly, upon the application of equation~(\ref{pressure.z}) in the $z$-momentum equation, it may be verified that
\begin{equation}
ikv_z = \left[\frac{\omega^2}{c^2k^2} - 1\right]^{-1} \partial_x v_x,
\end{equation}
leading to the relation
\begin{equation}
\partial_x v_x + ik v_z = \eta(x) \partial_x v_x,
\end{equation}
where,
\begin{equation}
\eta(x) = \left[1 - \frac{c^2 k^2}{\omega^2}\right]^{-1}.
\end{equation}
Upon further manipulation, a second-order differential equation for the eigenfunction $v_x$ may be obtained:
\begin{equation}
\partial^2_x v_x + \Theta(x) \partial_x v_x +\Phi(x) v_x = 0,\label{eq.eigfunc}
\end{equation}
where,
\begin{eqnarray}
\Theta(x) &=& \frac{2 - \Gamma_1 \eta^2 }{\Gamma_1 p_0\eta + B^2_0} \db,\\
\Phi(x) &=& \frac{\rho_0 \omega^2 - k^2 B^2_0}{\Gamma_1 p_0\eta + B^2_0}.
\end{eqnarray}
Equation~(\ref{eq.eigfunc}) was solved using the MATLAB boundary value problem solver bvp4c. The boundary conditions were chosen to be $v_x|_{x=0,L} = 0$, with the additional condition $\partial_x v_x|_{x=0} = 1$ required to solve for the eigenvalue $\omega$. Because of the linearity of the problem, there is no loss of generality due to this third condition. We show a sample eigenfunction calculation in Figure~\ref{eigfunc.examp} for the resonant mode with $\nu = 5.09$ mHz; theory and simulation show good agreement. The background magnetic field was chosen to be $B_0 = \tilde{b} \sqrt{2x}$, $\tilde{b} = 71.5~{\rm G}~{\rm Mm}^{-1/2}$, $\Gamma_1 = 1.5$, $p_0 = 1.21 \times 10^5 - B_0^2/2~ {\rm dyne~cm^{-2}}, \rho_0 = 2.78 \times 10^{-7} ~{\rm g~cm^{-3}}$, with $x$ expressed in Mm.

\acknowledgements
This work was possible with funding from grant HMI NAS5-02139. Thanks to Keiji Hayashi for the instructive discussions relating to the CT way of dealing with magnetic fields. Also, thanks to Robert Cameron, Ashley Crouch, Elena Khomenko, Tom Bogdan, and Tom Duvall, Jr. for many useful conversations.


\begin{thebibliography}{}
\bibitem[Abbett(2007)]{abbett} Abbett, W. P. 2007, \apj, 665, 1469
\bibitem[Balthasar et al.(1998)]{balthasar} Balthasar, H. et al. 1998, \solphys, 182, 65
\bibitem[Beli\"{e}n et al.(2002)]{belien} Beli\"{e}n, A. J. C. et al. 2002, Journal of Computational Physics, 182, 91
\bibitem[Berland et al.(2006)]{berland} Berland, J. et al. 2006, Computers and Fluids, 35, 1459
\bibitem[Bogdan et al.(1993)]{bogdan93} Bogdan, T. J., Brown, T. M., Lites, B. W., \& Thomas, J. H. 1993, ApJ, 406, 723
\bibitem[Bogdan \& Cally(1995)]{bogdan95} Bogdan, T. J., \& Cally, P. S. 1995, ApJ, 453, 919
\bibitem[Bogdan et al.(1996)]{bogdan96} Bogdan, T. J., Hindman, B. W., Cally, P. S., \& Charbonneau, P. 1996, ApJ, 465, 406
\bibitem[Bogdan(2000)]{bogdan} Bogdan, T. J. 2000, \solphys, 192, 373
\bibitem[Braun et al.(1992)]{braun92} Braun, D. C. et al. 1992, ApJ, 392, 739
\bibitem[Braun(1995)]{braun95} Braun, D. C. 1995, ApJ, 451, 859
\bibitem[Braun et al.(1987)]{braun87} Braun, D. C., Duvall, T. L., Jr., \& LaBonte, B. J. 1987, ApJ, 319, L27
\bibitem[Braun \& Birch(2006)]{aaron} Braun, D. C. \& Birch, A. C. 2006, ApJ, 647, L187
\bibitem[Brown et al.(1992)]{brown92} Brown, T. M. et al. 1992, ApJ, 394, L65
\bibitem[Cally(1995)]{cally95} Cally, P. S. 1995, \apj, 451, 372
\bibitem[Cally(2000)]{cally00} Cally, P. S. 2000, \solphys, 192, 395
\bibitem[Cally \& Bogdan(1997)]{cally97} Cally, P. S. \& Bogdan, T. J. 1997, \apj, 486, L67
\bibitem[Cameron, Gizon, \& Daiffallah(2007)]{cameron07} Cameron R., Gizon, L., \& Daiffallah, K. 2007, Astronomische Nachrichten, 328, 313
\bibitem[Cattaneo(1999)]{cattaneo} Cattaneo, F. 1999, \apj, 515, L39
\bibitem[Cheung et al.(2006)]{cheung} Cheung, M. C. M., Moreno-Insertis, F., \& Sch\"{u}ssler, M. 2006, 451, 303
\bibitem[Christensen-Dalsgaard et al.(1996)]{jcd} Christensen-Dalsgaard, J., et al. 1996, Science, 272, 1286 
\bibitem[Couvidat, Birch, \& Kosovichev(2006)]{couvidat} Couvidat, S., Birch, A. C., \& Kosovichev, A. G. 2006, ApJ, 640, 516
\bibitem[Crouch \& Cally(2003)]{crouch03} Crouch, A. D. \& Cally, P. S. 2003, \solphys, 214, 201
\bibitem[Donea et al.(2000)]{donea} Donea, A.~ C., Lindsey, C., \& Braun, D. C. 2000, \solphys, 192, 321
\bibitem[Duvall et al.(1993)]{duvall} Duvall, T. L., Jefferies, S. M., Harvey, J. W., \& Pomerantz, M. A. 1993, \nat, 362, 430
\bibitem[Duvall et al.(1996)]{duvall96} Duvall, T. L., Jr., D'Silva, S., Jefferies, S. M., Harvey, J. W., \& Schou, J. 1996, Nature, 379, 235
\bibitem[Duvall, Birch, \& Gizon(2006)]{duvall06} Duvall, T. L., Jr., Birch, A. C., \& Gizon, L. 2006, ApJ, 646, 553
\bibitem[Evans \& Hawley(1988)]{evans} Evans, C. R. \& Hawley, J. F. 1988, \apj, 332, 659
\bibitem[Gizon \& Birch(2002)]{gizon02} Gizon, L., \& Birch, A. C. 2002, ApJ, 571, 966
\bibitem[Gizon, Hanasoge, \& Birch(2006)]{gizon06} Gizon, L., Hanasoge, S. M., \& Birch, A. C. 2006, ApJ, 643, 549
\bibitem[Gizon \& Birch(2005)]{gizon05} Gizon, L. \& Birch, A. C. 2005, Living Reviews in Solar Physics, 2, 6
\bibitem[Hanasoge et al.(2006)]{hanasoge1} Hanasoge, S. M. et al. 2006, \apj, 648, 1268
\bibitem[Hanasoge \& Duvall(2007)]{hanasoge2} Hanasoge, S. M. \& Duvall, T. L., Jr. 2007, Astronomische Nachrichten, 328, 319
\bibitem[Hanasoge et al.(2007a)]{hanasoge3} Hanasoge, S. M., Duvall, T. L., Jr., \& Couvidat, S. 2007a, \apj, 664, 1234
\bibitem[Hanasoge et al.(2007b)]{hanasoge4} Hanasoge, S. M., Couvidat, S., Rajaguru, S. P., \& Birch, A. C. 2007b, \apj, {\it accepted}, arXiv, 0707.1369H
\bibitem[Hanasoge(2007)]{hanasoge.phd} Hanasoge, S. M. 2007, Ph. D. thesis, Stanford University, http://soi.stanford.edu/papers/dissertations/hanasoge/
\bibitem[Hanasoge et al.(2007c)]{hanasoge5} Hanasoge, S. M., Birch, A. C., Bogdan, T. J., \& Gizon, L. 2007c, \apj, {\it submitted}
\bibitem[Hindman et al.(1997)]{hindman97} Hindman, B. W., Jain, R., \& Zweibel, E. 1997, \apj, 476, 392
\bibitem[Hindman \& Brown(1998)]{hindman98} Hindman, B. W. \& Brown, T. M. 1998, ApJ, 504, 1029
\bibitem[Khomenko, Collados, \& Felipe(2007)]{khomenko} Khomenko, E., Collados, M., \& Felipe, T. 2007, arXiv, 0710.3335
\bibitem[Kosovichev \& Duvall(1997)]{kosovichev} Kosovichev, A. G. \& Duvall, T. L., Jr. 1997, SCORe proceedings, Ed.: F.P. Pijpers, J. Christensen-Dalsgaard, and C.S. Rosenthal, Kluwer Academic Publishers, 241
\bibitem[Lele(1992)]{lele} Lele, S. K. 1992, Journal of Computational Physics, 103, 16
\bibitem[Lindsey \& Braun(2005)]{lindsey} Lindsey, C. \& Braun, D. C. 2005, \apj, 620, 1107
\bibitem[Lites et al.(1982)]{lites} Lites, B. W., White, O. R., \& Packman, D. 1981, \apj,253, L386
\bibitem[Livingston et al.(2006)]{living} Livingston, W. et al. 2006, \solphys, 239, 41
\bibitem[Moretti et al.(2007)]{moretti} Moretti et al. 2007, \aap, 471, 961
\bibitem[Nagashima et al.(2007)]{nagashima} Nagashima, K. et al. 2007, PASJ, 59, S631
\bibitem[Pizzo(1990)]{pizzo} Pizzo, V. J. 1990, \apj, 365, 764
\bibitem[Pneuman, Solanki, \& Stenflo(1986)]{pneuman} Pneuman, G. W., Solanki, S. K., \& Stenflo, J. O. 1986, A \& A, 154, 231
\bibitem[Parchevsky \& Kosovichev(2007a)]{parchevsky} Parchevsky, K. \& Kosovichev, A. G. 2007a, \apj, 666, 547
\bibitem[Parchevsky \& Kosovichev(2007b)]{parchevsky007} Parchevsky, K. \& Kosovichev, A. G. 2007b, \apj, 666, L53
\bibitem[Rosenthal \& Julien(2000)]{rosenthal} Rosenthal, C. S. \& Julien, K. A. 2000, 532, 1230
\bibitem[Scherrer et al.(1995)]{scherrer} Scherrer et al. 1995, \solphys, 162, 129
\bibitem[Sch\"{u}ssler \& Rempel(2005)]{schussler} Sch\"{u}ssler, M. \& Rempel, M. 2005, \aap, 441, 337S
\bibitem[Shelyag et al.(2007)]{shelyag} Shelyag, S. et al. 2007, Astroph, 3076S
\bibitem[Schl\"{u}ter \& Temesv\`{a}ry(1958)]{schluter} Schl\"{u}ter, A. \& Temesv\`{a}ry, S. 1958, IAUS, 6, 236
\bibitem[Schunker et al.(2003)]{schunker} Schunker, H., Braun, D. C., Cally, P. S., \& Lindsey, C. 2005, ApJ, 621, 149
\bibitem[Thompson(1990)]{thompson} Thompson, K. W. 1990, Journal of Computational Physics, 89, 439
\bibitem[T\'{o}th(2000)]{toth} T\'{o}th, G. 2000, Journal of Computational Physics, 161, 605
\bibitem[Zhao \& Kosovichev(2006)]{zhao} Zhao, J. \& Kosovichev, A. G. 2006, ApJ, 643, 1317
\end{thebibliography}
\end{document}